\documentclass[9pt,twocolumn,twoside]{opticajnl}
\journal{opticajournal} 

\setboolean{shortarticle}{false}

\usepackage{graphicx}
\usepackage{fancyhdr}
\usepackage{subcaption}
\usepackage{caption}
\usepackage{siunitx}

\usepackage{lineno}

\title{Measuring the dipole phase of Bloch-trajectory harmonics using a monolithic interferometer}

\author[1,2]{Mohanad Awad}
\author[1,2]{Julian Späthe}
\author[1] {Paramjit C. Yadav}
\author[1] {Mandakini Bhat}
\author[3] {Zsolt Divéki}
\author[3,4] {Eric Cormier}
\author[3] {Bálint Kiss}
\author[5] {Eszter Baradács}
\author[5] {János Tomán}
\author[5] {Zolt\'{a}n Erd\'{e}lyi}
\author[6,7,8] {Thomas Siefke}
\author[1,2] {Adrian N. Pfeiffer}
\author[1,2,*] {Matthias K\"ubel}

\affil[1] {Institute of Optics and Quantum Electronics, Friedrich Schiller University, Jena, Germany }
\affil[2] {Abbe Center of Photonics, Friedrich Schiller University, Jena, Germany} 
\affil[3] {ELI-ALPS, ELI-HU Non-Profit Ltd., Wolfgang Sandner utca 3., Szeged, H-6728, Hungary} 
\affil[4] {Laboratoire Photonique Num\'erique et Nanosciences (LP2N), UMR 5298, CNRS-IOGS-Universit\'e Bordeaux, 33400 Talence, France}
\affil[5] {Department of Solid State Physics, Faculty of Science and Technology, University of Debrecen, P.O. Box 400, 4002, Debrecen, Hungary}
\affil[6] {Institute of Applied Physics, Friedrich Schiller University Jena, 07745 Jena, Germany}
\affil[7] {Fraunhofer Institute for Applied Optics and Precision Engineering, 07745 Jena, Germany}
\affil[8] {Ernst-Abbe-Hochschule Jena, 07745 Jena, Germany}
\affil[*] {matthias.kuebel@uni-jena.de}

\begin{abstract}
Uncovering the dipole phase of gas-phase high harmonic generation was instrumental to understanding the recollision physics underlying attosecond pulse generation. Corresponding measurements in the condensed phase have not yet yielded a consistent picture. Here, we present a compact and inherently stable approach to high-harmonic interferometry in thin-film solids. We employ it to reveal the dipole phase of high-harmonic generation in polycrystalline ZnO, driven by broadband mid-IR laser pulses. We demonstrate that, under the conditions of our experiments, recollisions facilitated by Bloch oscillations represent the dominant contribution to high-harmonic generation just above the bandgap.
\end{abstract}

\setboolean{displaycopyright}{false} 

\begin{document}

\maketitle

\section{Introduction}
High-harmonic generation (HHG) in solids has emerged as a sensitive and ultrafast probe of material properties and dynamics in bulk crystals and 2D materials \cite{goulielmakis2022high, heide2024ultrafast, Ciappina2025}. The sensitivity of HHG to electronic structure and ultrafast dynamics results from the coherent driving of charge carriers through the entire Brillouin zone within fractions of an optical cycle \cite{yue2021expanded,Reisloehner2022}, giving rise to Bloch oscillations \cite{schubert2014sub,wu2015high}. The rich quantum dynamics involved in solid HHG have allowed for the measurement of band structure \cite{Vampa2015a,hohenleutner2015real}, probing Berry curvature \cite{Luu2018, Uzan-Narovlansky2024}, revealing valley polarization \cite{liu2017high,yoshikawa2019interband,mrudul2021light}, and tracking lattice dynamics \cite{Zhang2024}. 
The HHG response also represents a sensitive probe of local field properties such as pulse duration \cite{Awad2024} and carrier-envelope phase (CEP) \cite{Hollinger2020}.

Accessing phase information through two-color driving has provided important insights into the mechanisms underlying HHG in solids \cite{Vampa2015,wang2023trajectory,Uzan-Narovlansky2024,Kim2025}. Crucially, Vampa, et al. showed that a three-step model, analogous to gas-phase HHG \cite{Corkum1993}, can explain features of HHG in solids \cite{Vampa2015,Vampa2015semi}. In addition, the so-called intraband mechanism, which does not invoke recombination but is governed by coherent Bloch oscillations, contributes to the high-harmonic spectrum \cite{Ghimire2011, Vampa2014}. The prevalence of different mechanisms has been shown to depend on the specific material and laser parameters \cite{schubert2014sub,Luu2015,tancogne2017ellipticity}.

Measuring the CEP dependence of HHG using single-cycle laser pulses, Garg et al. showed that HHG generated in solids is much less affected by intensity variations than gas-phase HHG \cite{Garg2018}. More recently, HHG experiments using extrinsic interferometers have been conducted \cite{lu2019interferometry, Uchida2024, koll2025extreme, Kuzkova2026}. Most interferometric experiments have utilized near-infrared (NIR) driving fields interacting with bulk dielectrics. The experimental results indicate that the dipole phase of solid HHG is indeed intensity dependent \cite{lu2019interferometry,Kuzkova2026}. Further, Koll et al. showed that the dipole phase in crystalline MgO and amorphous SiO$_2$ exhibits opposite behavior with increasing intensity \cite{koll2025extreme}. The variety of findings indicate that further measurements are required to obtain a comprehensive picture of HHG in solids.    


Here, we introduce a platform for HHG interferometry using semiconductor thin films in a common-path configuration. We employ it to measure the intensity-dependence of the HHG dipole phase for polycrystalline ZnO thin-films, driven by few-cycle mid-infrared (MIR) pulses. Our results provide evidence that harmonic just above the bandgap are produced by electron-hole recollisions, which are facilitated by Bloch oscillations.
In our implementation of high-harmonic interferometry, two sources are longitudinally separated along the laser propagation by a dispersive element, such as a dielectric substrate. A single mid-IR laser pulse is used to generate HHG in both sources, and the resulting signal interferes on the detector. 

\section{Methods}
\paragraph{Sample design:} The HHG targets consist of a sapphire substrate of \SI{500}{\micro m} thickness, coated with ZnO thin films. The thin films were grown by atomic layer deposition (ALD, Beneq TFS-200) at \SI{125}{\celsius} using diethylzinc (DEZ) and water as precursors. The precursor pulse durations were \SI{150}{ms}, followed by purge times of \SI{2.5}{s} (after DEZ) and \SI{51}{s} (after H$_2$O). During deposition, the chamber and reactor pressures were maintained at \SI{7.4}{mbar} and \SI{1.2}{mbar}, respectively. The ZnO thickness was obtained using 350 ALD cycles, resulting in \SI{47}{nm} thick layers, measured by spectroscopic ellipsometry. Both single- and double-sided coated substrates were prepared. When a sample is placed at the laser focus, high harmonics are generated in the ZnO layers. Measurements with blank substrates confirm that no high harmonics (of order $n\geq 5$) are generated in the substrate. We choose sapphire for its large transmission window, which allows the mid-IR fundamental and harmonics down to $\approx \SI{200}{nm}$ to be transmitted. Thus, we realize an ultra-compact HHG interferometer by coating the dielectric substrates with semiconductor layers on both sides. The thickness of the layers is \SI{50}{nm}, well below the wavelengths of the generated harmonics, and below the coherence lengths of typically a few \SI{}{\micro m} \cite{journigan2024high}. The absorption length for above-gap harmonics, i.e., $\lambda<\SI{380}{nm}$ is approximately \SI{40}{nm}, similar to the layer thickness. Otherwise, propagation effects within the thin-films are considered insignificant.

\paragraph{HHG experiments:} HHG experiments are conducted using the mid-infrared (MIR) laser at ELI ALPS, which provides ultrashort pulses with a duration of \SI{42}{fs}, centered at a wavelength of $\lambda_1 = \SI{3.2}{\micro m}$. The laser pulses are spectrally broadened via self-phase modulation (SPM) induced by nonlinear propagation through BaF$_2$ and Si windows. 
Subsequent chirp compensation is achieved by linear propagation through CaF$_2$ and BaF$_2$ bulks, while third order phase is corrected by adequate multilayer mirrors resulting in \SI{24}{fs} compressed pulses \cite{Kurucz2020}. The laser polarization and pulse energy are controlled using an achromatic half-wave plate followed by a pair of broadband wiregrid polarizers. Pulses with an energy of the order of \SI{10}{\micro J} are focused ($f = \SI{20}{cm}$) with a spot size of $\approx \SI{100}{\micro m}$ (M$^2 = 1.6$) onto a solid target to drive high-harmonic generation. The spectra of the emitted harmonics are measured using a compact UV-VIS spectrometer (Ocean Fx). 

\paragraph{Pulse propagation simulations:} Experimental results are compared to simulations based on the numerical solutions of the one-dimensional forward Maxwell equation (FME) in a co-moving reference frame \cite{Husakou2001},

\begin{align}
    \frac{\partial \tilde{E}(\omega, z)}{\partial z} = &-i \left( k_z(\omega) - k_{\text{ref}}(\omega) \right) \tilde{E}(\omega, z) \\
    &- i \frac{\omega}{2 \epsilon_0 c n(\omega)} \tilde{P}_{\text{NL}}(\omega, z),
\end{align}

where $\tilde{E}(\omega, z)$ is the complex electric field in the frequency domain, and $k_z(\omega) = \frac{\omega}{c} n(\omega)$ is the longitudinal wavevector component, which incorporates the linear dispersion derived from empirical Sellmeier equations. 
$k_{\text{ref}}(\omega) = k(\omega_0) + \frac{\omega - \omega_0}{v_g}$ is the reference wavevector. The source term for third-order nonlinearities is evaluated in the time domain as 
\begin{equation}
P_{\text{NL}}(t, z) = \epsilon_0 \chi^{(3)} \vert{}E(t, z)\vert{}^2 E(t, z),    
\end{equation}
and Fourier transformed at each spatial step to yield $\tilde{P}_{\text{NL}}(\omega, z)$.

Calculations are carried out for a chirp-free laser pulse with a Gaussian intensity envelope of \SI{25}{fs} duration (full width at half maximum), centered at a carrier wavelength of \SI{3200}{nm}.  The variation of the laser intensity and phase throughout the focus is described according to Gaussian beam optics. The microscopic high-harmonic emission is treated phenomenologically, with the field of harmonic order $q$ scaling dynamically with the fundamental local driving field as
\begin{equation}
 E_{q} = \left|E_{\text{local}}\right|^{N_{\text{eff}}} \exp{\left(i (q \phi -\phi_\mathrm{HHG}\right)}, 
\end{equation}
where $\phi$ is the phase of the fundamental and $\phi_\mathrm{HHG}$ is the dipole phase of HHG.
The same empirical effective non-linearity $N_\text{eff} = 2.5$ is used for all harmonic orders. To obtain the local electric field $E_{\text{local}}$, the transmission of the fundamental and harmonics at each interface is treated analytically, using complex Fresnel coefficients. The model explicitly evaluates the Fabry-Perot interference resulting from multiple internal reflections within the thin film. Thus, HHG is driven by the locally enhanced internal electric field rather than the vacuum intensity. 

The propagation of the light field, consisting of fundamental and front-side harmonics, through the dielectric substrate is evaluated using a fourth-order Runge-Kutta integration scheme in a moving reference frame centered on the pump group velocity. The propagation accounts for linear dispersion—derived from empirical Sellmeier equations, as well as for third-order nonlinearities (self-phase modulation and cross-phase modulation). Although the peak power of the fundamental pulse exceeds the critical power for self-focusing in sapphire, the $\SI{500}{\micro m}$ substrate thickness is considerably shorter than the estimated nonlinear collapse distance ($z_f \geq \SI{1}{mm}$), and the total accumulated B-integral (estimated at the peak intensity of $\SI{6}{TW/cm^2}$ used in the experiment) remains sub-critical ($B < \pi$). This justifies the use of a 1D pulse propagation model.

\paragraph{Calculation of the high-harmonic dipole phase}
We calculate the dipole phase of HHG using the semi-classical theory of HHG in solids \cite{Vampa2015semi}. This model assumed that HHG takes place via the recollision mechanism, in close analogy to gas-phase HHG. Following excitation at $t_i$, the electron and hole are propagated in the conduction and valence bands, according to 
\begin{equation}
    v\left[\mathbf{k}(\tau)\right] = v\left[\mathbf{A}(\tau)-\mathbf{A}(t_i)\right],
\end{equation}
where $v(\mathbf{k}) = \nabla_k \Delta E(\mathbf{k})$ is the group velocity corresponding to the relative position-space motion of electron and hole. Here, $\Delta E$ is the energy difference of electron and hole, corresponding to the band structure. In the light of the polycrystalline samples used in our experiment, we approximate the band structure of ZnO using sinusoidal bands with a bandgap of \SI{3.2}{eV} and a bandwidth of \SI{5.0}{eV}. If electron and hole re-encounter each other at a time $t_f$ within a single optical cycle, they may recombine and a photon corresponding to the energy difference $E(t_f)$ is emitted. Associated with the electron-hole trajectory is a phase corresponding to the classical action accumulated during the propagation of the electron-hole pair. Taking into account the propagation of the emitted photon in the continuum, the dipole phase of harmonic $q$ can be expressed as \cite{Kuzkova2026}
\begin{equation}
    \phi_q = q\left(\omega_0 t_f + \frac{\pi}{2}\right) - \int_{t_i}^{t_f} \Delta E[\mathbf{k}(\tau)] \text{d}\tau.
\end{equation}

\section{Results and discussion}
\subsection{High harmonic spectra}
\begin{figure}[h]
    \centering
    \includegraphics[width=\linewidth]{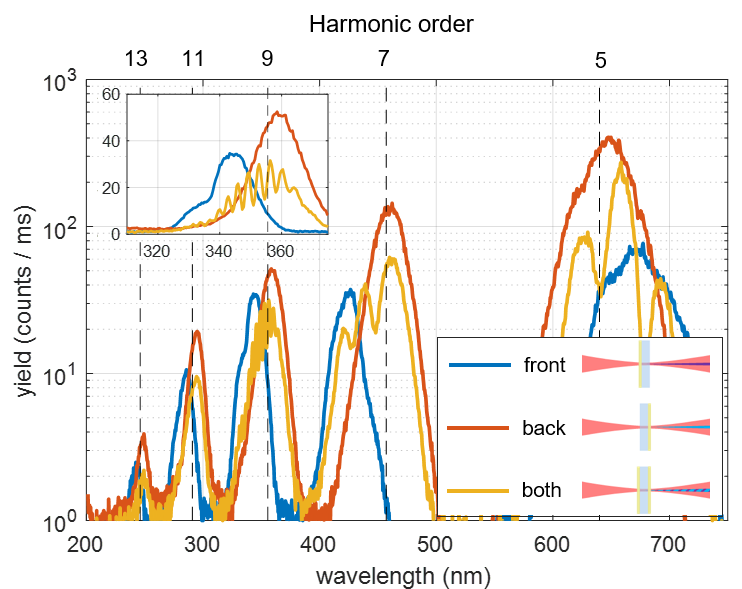}
    \caption{Typical HHG spectra recorded for single-sided and double-sided samples consisting of 50 nm ZnO films deposited on sapphire. The legend depicts all three geometries used in the present experiment. The inset zooms into the region around the 9th harmonic, where narrow interference fringes are observed for the double-sided samples.}
    \label{fig1}
\end{figure}

Fig.~\ref{fig1} shows typical HHG spectra measured for single and double-sided ZnO samples. For the single-sided samples, we distinguish two cases: first, the \emph{front} geometry, where the ZnO layer is facing the incident laser beam such that harmonics are generated in the front of the sample; and second, the \emph{back} geometry, where the laser pulse propagates through the substrate before generating harmonics in the ZnO layer at the back of the sample. As seen in Fig.\ref{fig1}, the HHG signal in the back geometry is significantly  stronger than in the front geometry. In all cases, harmonics up to order 13 (in short H13) are recorded in the present experiments. Interestingly, the front and back harmonic spectra differ significantly. The front-side harmonics of order $n \geq 7$ are blue-shifted with respect to the back-side harmonics, and also with respect to the expectation $\lambda_1 / n$, as indicated by the dashed lines. In contrast, the 5th harmonic generated in front-geometry is red-shifted.

The HHG spectra recorded for the double-sided samples exhibit broad peaks with the center situated in-between the central wavelengths observed for front and back geometry in the single-sided samples. For H5, H7 and H9, clear interference fringes are observed. The visibility of the fringes is below unity, owing to the weaker harmonic emission from the front side and partial absorption of the front harmonics in the back ZnO layer.
For short wavelengths, the spacing between the fringes becomes increasingly narrow. For H11 and H13, no fringes are observed. Our calculations detailed below show that the expected fringe spacing in the H11 signal is \SI{1}{nm}. Thus, the spectral resolution of the compact spectrometer ($\delta \lambda \approx \SI{1}{nm}$) used in our experiment is insufficient to resolve the fringes below \SI{300}{nm}.


\subsection{Measurement of linear dispersion: coherence tomography}
\begin{figure*}
    \centering
    \includegraphics[width=0.8\linewidth]{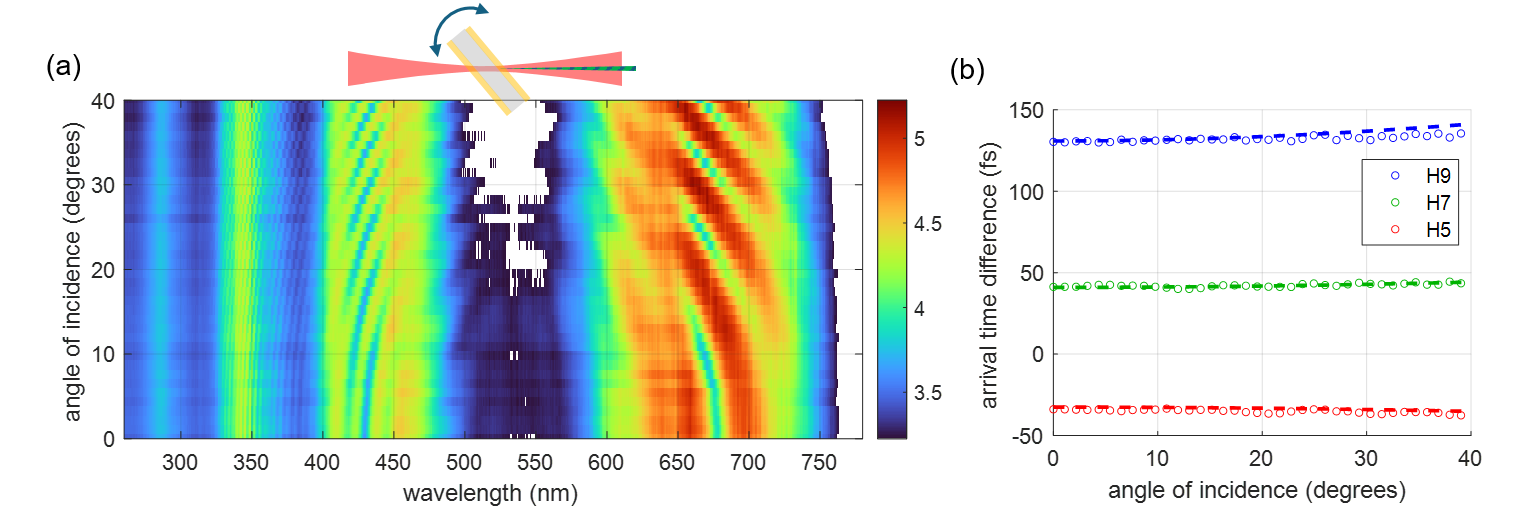}
    \caption{Tomographic HHG scan: (a) Series of HHG spectra measured as a function of angle of incidence (AOI) for a double-layered sample ($\SI{50}{nm}$ ZnO on a $\SI{500}{\micro m}$ thick sapphire substrate. (b) Retrieved arrival time difference of harmonics emitted from the front and back surfaces, as a function of the angle of incidence. The dashed lines show the calculated arrival time differences for H5, H7 and H9 assuming linear dispersion only. Errorbars are omitted as they would be smaller than the size of the data points.}
    \label{fig2}
\end{figure*}

For an experimental verification of our interferometric approach, we vary the optical pathlength in the interferometer by rotating the sample surface with respect to the laser propagation direction. For each rotation angle, HHG spectra are recorded, resulting in the tomographic interferogram presented in Fig.~\ref{fig2}(a). With increasing angle of incidence (AOI), the fringes observed for H5 shift in the opposite direction as those observed for H7 and H9. For H9, the pixel resolution of the spectrometer reaches its limits for AOI greater than approximately $30^\circ$.

The fringe spacing in the interferogram is given by the arrival time difference $\delta t$ between the harmonics generated in the front and in the back by 
\begin{equation}
\delta \lambda  = \frac{\lambda^2}{c \delta t}. 
\end{equation}
We retrieve the arrival time difference by taking a Fourier transform of the measured spectra. The resulting arrival time difference for H5, H7 and H9 is plotted at each angle of incidence in Fig.~\ref{fig2}(b). The measured delays are compared to calculated ones, based on the linear refractive indices of the fundamental and the harmonics. The calculations agree  well with the the measured phase difference. The opposite sign of the phase difference observed for H5 on one hand and H7 and H9 on the other hand reflect the group velocity dispersion; at $\SI{530}{nm}$, the group velocity coincides with that of the infrared fundamental centered at $\SI{3200}{nm}$. The deviation of the measured arrival time difference from the calculated value observed for H9 at large AOI is attributed to the limited spectral resolution. 

\subsection{Intensity dependence of interferometric HHG spectra}
\begin{figure*}
    \centering
    \includegraphics[width= 0.9\linewidth]{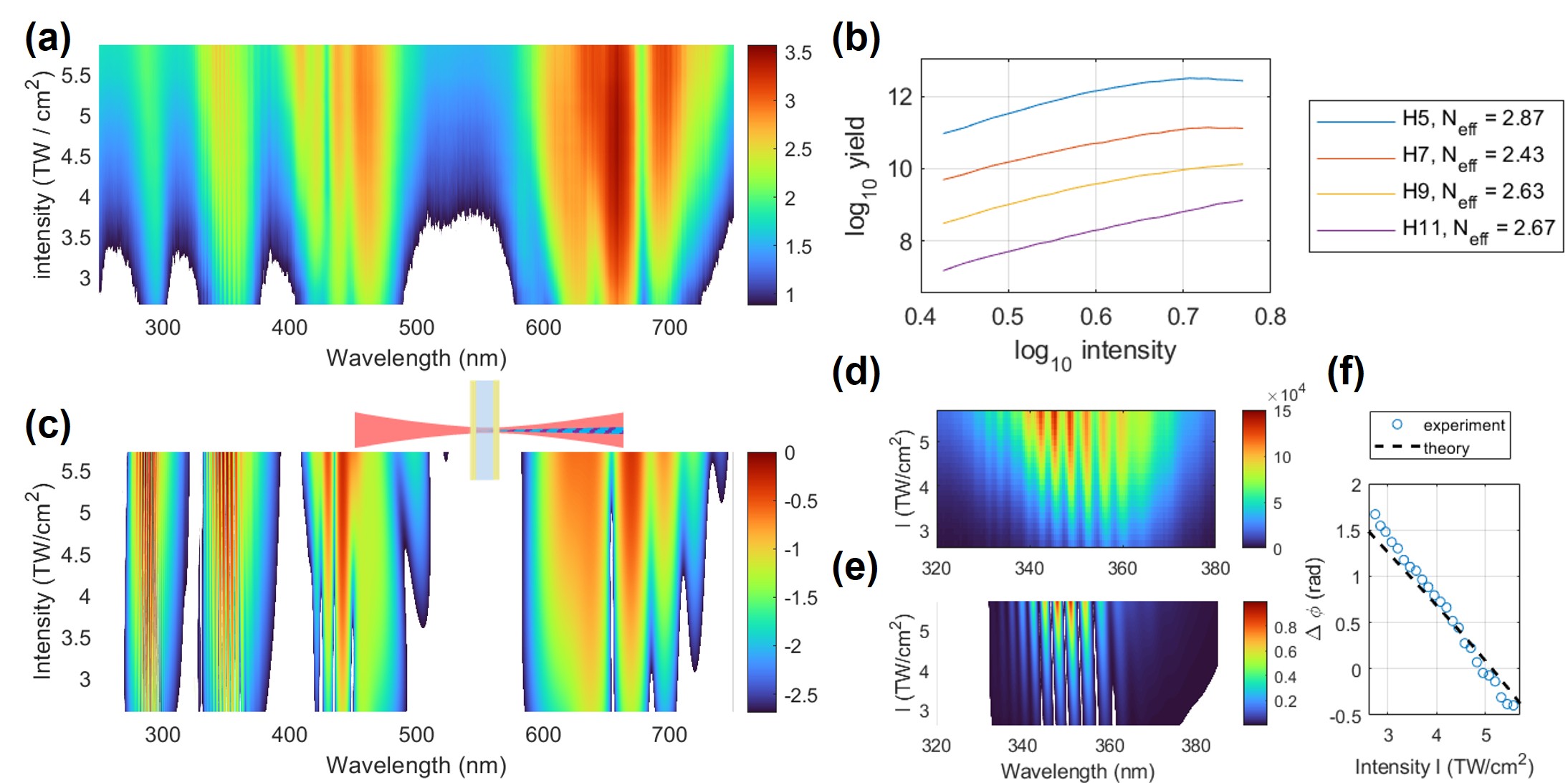}
    \caption{Non-linear behavior of the high-harmonic interferometer. Shown is the measured HHG yield as a function of (a) wavelength and peak intensity in vacuum. The effective non-linearity $N_\text{eff}$ for each harmonic order is obtained by fitting a first-order polynomial to the measured intensity-dependent yield (b). Panel (c) shows the computational results for the intensity-dependent high-harmonic interferograms using $N_\text{eff} = 2.5$ for all orders. The measured (d) and calculated (e) intensity-dependent signals for the 9th harmonic signal are analyzed for the phase encoded in the interferogram, which is displayed as a function of intensity in (f).}
    \label{fig:intensity_scan}
\end{figure*}

In order to understand the different responses observed for front and back geometries (see Fig.~\ref{fig1}), and for the accurate characterization of the interferometric samples in general, we perform measurements of the intensity dependence of HHG in single- and double-sided samples. The laser intensity is varied by rotating the polarizer while keeping the analyzer fixed to allow s-polarized light to pass.
The experimental results for the double-sided samples are displayed in Fig.~\ref{fig:intensity_scan} (a-c) along with computational results, obtained using the pulse propagation model. 
The computational results agree very well with the experimental ones. This allows us to analyze the physics underlying our observations; namely, the differences between the spectra of front-side and back-side harmonic emission, as well as the enhanced yield observed for the back geometry. 


The harmonics generated in the front layer co-propagate with the intense fundamental through the substrate. The presence of the intense fundamental modifies the refractive index of the material by the optical Kerr effect. As a consequence, the spectra of the front-side harmonics are affected by cross-phase modulation. Because the group velocity of H5 is higher than that of the fundamental, H5 propagates on the rising slope of the fundamental pulse, where the intensity of the fundamental pulse increases as a function of time. Due to the Kerr effect, the refractive index also increases as a function of time, $\frac{\partial n}{\partial t}>0$, leading to a red-shift of H5. The harmonics of order 7 and higher, which have a lower group velocity than the fundamental, experience the falling slope of the fundamental pulse, resulting in a blue shift of the harmonics. H7 acquires only $\approx \SI{35}{fs}$ group delay with respect to the fundamental over the entire $\SI{500}{\micro m}$ thickness of the substrate. Hence, it remains at the falling slope of the intense fundamental essentially throughout the entire substrate, resulting in a pronounced blue shift. Higher harmonic orders are quickly outrun by the fundamental pulse due to the increasingly large group velocity mismatch with decreasing wavelength, leading to a lesser blue shift.

The harmonics generated in the back of the sample are not affected by cross-phase modulation but inherit the effect of self-phase modulation of the fundamental pulse. We observe a slight spectral broadening of the fundamental, leading to a decrease of the Fourier transform limit from \SI{20.0}{fs} to \SI{17.5}{fs} after passing through a sample placed at the focus position. The spectral broadening of the fundamental results in the spectral broadening of all harmonic orders. The broadening can lead to the spectral overlap of successive harmonic orders, as can be seen in the experimental data around 400 nm. The resulting interference fringes are sensitive to the CEP of the driving laser \cite{Hollinger2020}. 

Our model also allows us to understand why the harmonic yield in the back geometry is stronger than in the front geometry: at the ZnO-vacuum interface at the back side of the sample, the constructive interference between the reflected and transmitted waves leads to local field enhancement. At the front side, the reflected and transmitted waves at the air-ZnO interface interfere destructively due to the phase shift at the optically dense medium. At the ZnO-sapphire interface, the reflectivity is small due to the similar refractive indices (respectively, 1.91 and 1.71 at \SI{3200}{nm}). In addition to these thin-film effects, non-linear effects contribute to the yield enhancement at the back side: our calculations show that the ratio of the local intensities at the front and back layers increases with increasing intensity. 

The signal measured for the double-sided samples exhibits intensity-dependent interference fringes. Using a Fourier transform, the relative phase encoded in the fringes can be retrieved. We have verified for the numerical results that the retrieved phases agree accurately with the true phase differences of the front-side and back-side emissions. This allows us to extract the relative phases of front-side and back-side HHG emissions from the experimental data.

We analyze the H9 interferograms for their encoded phase, as it exhibits several interference fringes, facilitating an accurate phase retrieval see Fig.\ref{fig:intensity_scan}(d-f). To extract the phase difference, the data is first transformed to wavenumber. Next, a Fourier transform is taken. The phase difference $\Delta \phi$ between the front-side and back-side harmonic emission is then extracted at the central modulation frequency and displayed in Fig.~\ref{fig:intensity_scan}(f) as a function of intensity. 
The computational results, obtained by using the literature value of $n_2 = \SI{2e-20}{m^2/W}$ at the fundamental wavelength, agree well with the measured intensity-dependent phase difference. However, the calculations underestimate the slope of the intensity-dependent phase shift by approximately 15\% compared to the measurement. This deviation could be due to an underestimation of the laser intensity or $n_2$.
Moreover, the intensity-dependent dipole phase of the HHG process itself can contribute to the interferometric delay when the laser intensities at the front and at the back of the samples differ. 

\subsection{Phase of the high harmonics}

The intensity-dependent HHG dipole phase is investigated by a combined scan of the pulse energy and sample position $z$ along the laser focus. Varying the intensity difference  between the front and back layers via the $z$ scan makes our measurement sensitive to the intrinsic dipole phase of HHG. The scan of the peak intensity using the polarizer allows us to remove uncertainties related to the non-linear pulse propagation in the substrate. 

\begin{figure*}
    \centering
    \includegraphics[width= \linewidth]{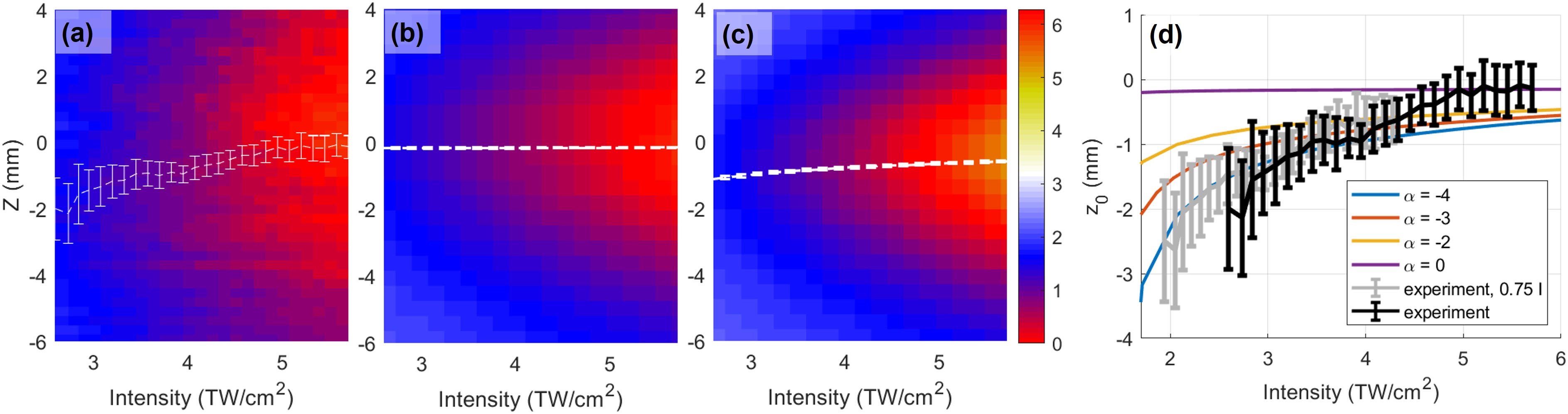}
    \caption{Combined intensity-z scans reveal the imprint of the HHG dipole phase. Shown is the phase difference in the 9th harmonic signal extracted from the (a) measured and (b,c) calculated interferograms. Calculations are performed (b) without and (c) with consideration of an intensity-dependent HHG dipole phase $\phi_\mathrm{HHG} = \eta I$. The black data points indicate the position $z_0$ at which the phase shift is maximized for each power value. The error bar indicate the uncertainty of the fit results. (d) The measured intensity dependence of $z_0$ is compared to computational results for various values of $\eta$.}
    \label{fig:intensity_z_scan}
\end{figure*}

HHG spectra are recorded for each sample position and each value of the pulse energy. The phase encoded in the interferogram of the 9th harmonic, as a function of peak intensity and sample position is displayed in Fig.~\ref{fig:intensity_z_scan}(a). As seen above, the phase depends on the the driving laser intensity. Additionally, the phase depends also on the sample position, and exhibits an apex at a position $z_0 (I)$ close to the center of the focus. Qualitatively, this behavior is consistent with the intuition that the nonlinear phase shift gets stronger with increasing intensity and is maximized when the sample is placed at the focus. In addition, the Gouy phase may contribute to the $z$-dependence of the phase shift, though the Rayleigh range $z_R \approx \SI{4}{mm}$ is much larger than the substrate thickness. 

The variation of $z_0$ with intensity deserves further investigation. We determine its value using a parabolic fit to the experimental results. Remarkably, $z_0(I)$ varies by $>\SI{1}{mm}$ within the scanned intensity range. Given the sub-critical value of the B-integral and the $\SI{0.5}{mm}$ thickness of the substrate, self-focusing cannot explain this strong variation of $z_0 (I)$. Furthermore, in the computational results assuming only Kerr nonlinearities in the substrate, Fig.~\ref{fig:intensity_z_scan}(b), $z_0$ does not depend on intensity. However, when an intensity-dependent dipole phase of HHG $\phi_\mathrm{HHG}(I) = \eta I$ is considered, $z_0$ becomes intensity dependent, as seen in Fig.~\ref{fig:intensity_z_scan}(c). 

The measured behavior of $z_0$ agrees with the computational results for $\eta \approx \SI{3}{\frac{rad cm^2}{TW}}$, as shown in Fig.~\ref{fig:intensity_z_scan}(d). The comparison suggests that the experimental intensity is somewhat overestimated. 
Notably, our experimental result for the intensity-dependence of the dipole phase is significantly larger than the values reported previously for HHG from dielectrics using NIR driving lasers \cite{lu2019interferometry,koll2025extreme,Kuzkova2026}. This raises the question about the underlying mechanism.

\begin{figure}
    \centering
    \includegraphics[width=0.9\linewidth]{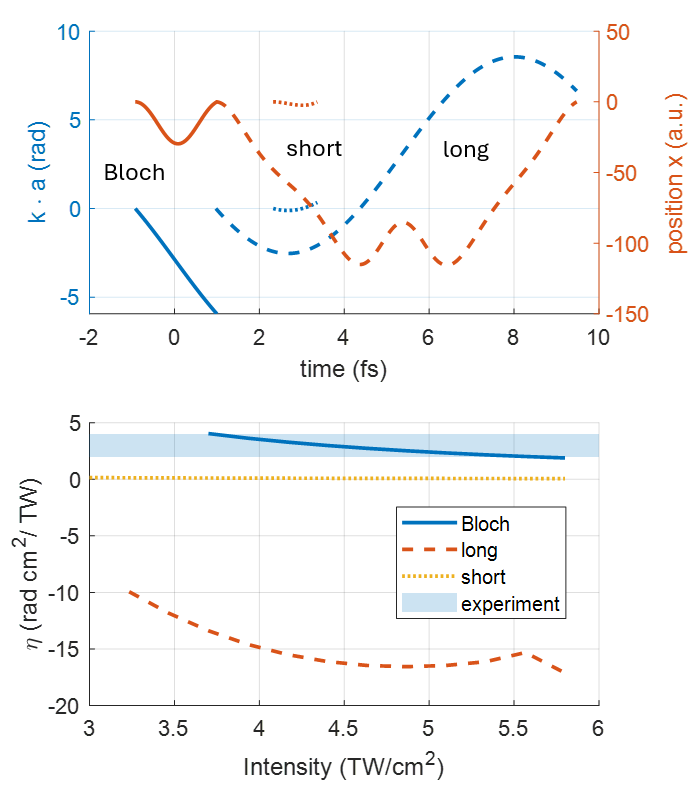}
    \caption{(a) Calculated electron-hole trajectory in k-space and position-space, leading to the emission of the 9th harmonic via three mechanisms as indicated in the figure. (b) The calculated intensity dependence of the dipole phase for the short, long and Bloch trajectories.}
    \label{fig:appendix:trajectory}
\end{figure}

We evaluate our experimental finding for the intensity-dependent dipole phase of the 9th harmonic by calculating the harmonic phase using the semi-classical recollision model described in the method section. The 9th harmonic energy of \SI{3.6}{eV} is only slightly above the bandgap of \SI{3.2}{eV}, corresponding to a recollision energy of only \SI{0.4}{eV}. Within the semi-classical treatment of HHG, low recollision energies correspond to either very short or very long trajectories, see dotted and dashed curves in Fig.~\ref{fig:appendix:trajectory}(a), respectively. The long trajectories possess an ionization time close to the peak of the field and travel times longer than $\approx 0.7 T $, where $T$ is the optical period. Short trajectories are born at a later time and possess shorter travel times. In the case of low recollision energies ($\approx\SI{0.4}{eV}$ for the 9th harmonic under the conditions of our experiment), both types of trajectories are expected to be suppressed for different reason. For the short trajectory $E(t_i)$, the field at the ionization time, and, hence, the ionization amplitude is low. For the long trajectories, ultrafast dephasing \cite{Vampa2015semi} suppresses their contribution due to the long travel time. 

The solid medium allows for recollisions via a third type of trajectories: Bloch oscillations. Using the semi-classical model, we find that these Bloch trajectories possess a large ionization amplitude while having a short travel time. Fig.~\ref{fig:appendix:trajectory}(a) displays a representative trajectory. It is initiated at $t_i \approx - 0.1 T$, i.e. before the peak of the field, and recollides at $t_r \approx 0.1 T$, briefly after the peak. On this trajectory, the crystal momentum never changes its sign during the propagation. However, the position space motion reverses its direction once the electron reaches the edge of the Brillouin zone, which occurs briefly after the peak of the field, i.e. after only $0.1 T$ of propagation. Within the following $\approx 0.1 T$ the electron-hole pair completes a Bloch oscillation. Fig.~\ref{fig:appendix:trajectory} (b) shows the calculated intensity dependence of the dipole phase for the Bloch oscillation trajectory. At an intensity around \SI{4}{\frac{TW}{cm^2}}, the intensity dependence of the dipole phase exhibits a slope of $\approx \SI{3.5}{\frac{rad \cdot cm^2}{TW}}$, which is in good agreement with the experimental results. In contrast, the calculated intensity dependence of the short (long) trajectories is much weaker (stronger).


\section{Conclusions and Outlook}
Our compact implementation of interferometry using high harmonics generated in thin-film solids is inherently stable, owing to its monolithic design. The scheme enables coherence tomography in transmission using harmonics within the transparent window of the dispersive material between the harmonic sources, i.e~$\lambda \geq \SI{250}{nm}$ for the sapphire substrates used in the present work. The spectra of the harmonics generated in the front and in the back are affected by cross-phase and self-phase modulation, which can be exploited to shape the harmonic spectra, increase their bandwidth and thereby optimize the phase resolution. We demonstrate that our interferometer can accurately resolve the non-linear phase accumulated in the dispersive material. Further, we present evidence that HHG in wide-bandgap semiconductors, driven by mid-IR lasers, exhibits a strongly intensity-dependent dipole phase due to Bloch oscillations. Our accessible interferometric scheme will be particularly useful as a platform to probe material-specific responses, effects of layer thickness and for time-resolved interferometry in non-collinear pump-probe experiments, where a pump pulse induces a phase shift in the harmonics generated on either side of the sample.

\section{Backmatter}

\begin{backmatter}
\bmsection{Funding} This project has been supported by the Deutsche Forschungsgemeinschaft (DFG, Germany Science Foundation) in the Collaborative Research Centre 1375 "Nonlinear optics down to atomic scales" (NOA) under projects B1 and Z3; and under the Emmy Noether programme, project No. 437321733. Also, ELI ALPS project (GINOP-2.3.6-15-2015-00001) is supported by the European Union and co-financed by the European Regional Development Fund. In addition, this work was supported by Project No. 2025-1.2.1-HU-RIZONT-2025-00006, implemented with the support provided from the National Research, Development and Innovation Fund of the Ministry of Culture and Innovation under the 2025-1.2.1-HU-RIZONT funding program and supported by the University of Debrecen Program for Scientific Publication.

\bmsection{Acknowledgments} We acknowledge fruitful discussions with P.M. Kraus, L. Faeyrman, D. Kartashov and S. Gräfe. We thank Thomas Weber and the technical staff at ELI ALPS for their support.

\bmsection{Disclosures} The authors declare no conflicts of interest.

\bmsection{Data Availability Statement} Data underlying the results presented in this paper are not publicly available at this time but may be obtained from the authors upon reasonable request.

\end{backmatter}

\section{References}

\bibliography{shhgi.bib}

\end{document}